\documentclass[11pt,tightenlines,superscriptaddress]{revtex4}
\usepackage{graphicx}
\usepackage{natbib}
\usepackage{url}

\newcommand{\CQT}{Centre~for~Quantum~Technologies, National~University~of~Singapore, 3 Science Drive 2, 117543, Singapore}
\newcommand{\Tsing}{Center~for~Quantum~Information, Institute~for~Interdisciplinary~Information~Sciences,\\ Tsinghua~University, Beijing, 100084, China}
\newcommand{\Oxf}{Atomic~and~Laser~Physics, University~of~Oxford, Clarendon~Laboratory,\\Parks Road, Oxford, OX1 3PU, United Kingdom.}

\begin{document}

\title{The ultimate physical limits to reversibility}
\author{Andrew J. P. Garner}
\email{ajpgarner@nus.edu.sg}
\affiliation{\CQT}
\affiliation{\Tsing}

\author{Vlatko Vedral}
\email{vlatko.vedral@qubit.org}
\affiliation{\Oxf}
\affiliation{\CQT}
\affiliation{Department of Physics, National University of Singapore, 2 Science Drive 3, Singapore 117542}
\affiliation{\Tsing}

\date{\today}

\begin{abstract}
We argue that if, in order to reverse an object's dynamics, we need to be able to keep track of it with enough precision, then there is an upper bound on the size of the object whose dynamics we can reverse -- even using all the available resources of the universe. 
Using a bound from the holographic principle and treating the universe as the ultimate quantum computer, we show that if we want to reverse the dynamics of an object which has evolved since the beginning of time, its radius cannot exceed ten microns.
If this computation is performed irreversibly, the object must be much smaller.
This places a lower bound on the size of the smallest possible quantum measurement device.  
\end{abstract}

\maketitle

Boltzmann's derivation of the H-theorem~\cite{Boltzmann1872}, proving the increase in entropy of an isolated gas whose molecules evolve by collisions dictated by Newtonian dynamics, was famously challenged by Loschmidt~\cite{Loschmidt1876}. s
Loschmidt argued that, if -- at some point -- the velocities of all the atoms were all to be reversed ($v$ to $-v$), the system would then evolve exactly back to its initial configuration. 
The manoeuvre where the velocities (or, in quantum physics, phases~\cite{Hahn50}) are reversed is now affectionately known as the Loschmidt echo. 
If in the forward direction the entropy increases, then in the backward direction it would have to decrease. 
With this simple thought experiment Loschmidt concluded that Boltzmann did not, and in fact could not, prove any entropy increase for a closed Newtonian system. 
Boltzmann is rumoured to have replied, ``Go ahead and reverse~them.''

This witty response is taken to imply that it is actually extremely hard to track molecular collisions in order to keep account of all the velocities and positions of all the molecules required to reverse them at some point during the evolution. 
But what exactly does ``extremely hard" mean and what kind of resources did Boltzmann actually have in mind? 

First of all, it is clear that if you are just given the state of, say, a mole of gas at some time and are asked to reverse all the velocities, this would require you to know all the positions of all the particles as well as their velocities (classically speaking). 
Then you could carefully position tiny walls which would bounce each of the particles so that the action exactly reverses their velocities. 
This kind of ability is akin to what Maxwell imagined when he invented his demon to (try to) violate the Second Law~\cite{Maxwell1870}. 
This may sound like a tall order, but what if the demon was to keep track of the evolution of the gas from the beginning and then at some point decide to act to reverse each collision? 
Keeping track of all the collisions might not be that hard, which was precisely Maxwell's point. 

Here, we quantify how hard it would be to keep track of the (simplified) dynamics of a system by treating the universe as the ultimate quantum computer and using all of its available resources -- space, time and energy -- to achieve this task. 
To derive the ultimate bound on the size of the system and the amount of time its dynamics could be followed and recorded faithfully, we appeal to the holographic principle (that the the information capacity of a volume of space is proportional to its surface area)~\cite{tHooft93,Susskind95,Bousso02} and view the universe as the ultimate quantum computer~\cite{Lloyd00,Lloyd02}. 
According to this, the maximum number of bits $I_S$ required to describe an object of radius $r_S$ is
\begin{equation}
I_S = \left(\frac{r_S}{l_P}\right)^{\!2},
\end{equation}
where $l_P$ is the Planck length $\sim 10^{-35}\; \mathrm{m}$~\cite{Planck1899,NIST}. 
Imagine that we need to keep track of this many bits for a duration $t$.
If we assume that time ticks with a period of Planck's time $t_P \sim 10^{-44}\; \mathrm{s}$~\cite{NIST}, then the bits describing this object could in principle change their state at this rate. 
The total amount of information $I_T$ is then 
\begin{equation}
I_T = \frac{t}{t_P} \left(\frac{r_S}{l_P}\right)^{\!2}.
\end{equation}
This amount must not exceed the total amount of information in our (visible) universe (radius $r_U$) which is equal to 
\begin{equation}
I_U = \left(\frac{r_U}{l_P}\right)^{\!2}
\end{equation}
from which we obtain
\begin{equation}
I_T = \frac{t}{t_P} \left(\frac{r_S}{l_P}\right)^{\!2} \leq I_U = \left(\frac{r_U}{l_P}\right)^2 .
\end{equation}
Solving for $r_S$ we obtain
\begin{equation}
r_S \leq r_U \sqrt{\frac{t_P}{t}}.
\end{equation}
Let us now take $t$ to be the present age of the universe ($\sim 13.8\; \mathrm{GYr}$~\cite{Planck15}). 
This leads to $r_S \leq {\sim 10^{-5}\;\mathrm{m}}$. 
The biggest object whose dynamics we can track to the maximum level of detail for the duration of the entire evolution of the universe thus far therefore has a size of the order of ten microns. 

Interestingly, we can arrive at the same conclusion by another computationally motivated argument. 
The energy required to flip a bit at the rate of $t_P$ is of order $h / 4 t_P = \; \sim 10^9~\mathrm{J}$~\cite{Lloyd00}.
The total energy in the universe can be taken to be $m_U c^2$ where the mass of the universe is taken to be $m_U = 10^{53}~\mathrm{kg}$~\cite{Valev14}.
The total number of bits we can therefore flip in Planck's time using the total available energy in the universe is therefore about $10^{60}$. By using the holographic argument, the size of the object with this many bits is 
\begin{equation}
10^{60} = \left(\frac{r_S}{l_P}\right)^{\!2},
\end{equation}
which gives us the same result of order ten microns. 

The arguments here assume that the tracking is done reversibly. 
Imagine that the bit flips are irreversible and require a resetting energy of $kT \ln 2$ per bit~\cite{Landauer61,Bennett82,LeffR02,MaruyamaNV09,BerutAPCDL12,ParrondoHS15}, where $T$ is assumed to be on the order of the cosmic microwave background temperature of the universe ($\sim 3\;\mathrm{K}$~\cite{Fixsen09}) and $k$ is Boltzmann's constant. 
Then, using all the energy of the universe, the largest number of bits we could possibly reset is $10^{92}$.
So if we reset at each Planck time step for the entire duration of the evolution of the universe thus far, this will mean that the number of such bits is $10^{33}$. 
From the Holographic bound, this corresponds to an object whose radius is about a tenth of an attometer ($10^{-19}\;\mathrm{m}$). 
As expected, the irreversible way of computing leads to much less efficient tracking, and therefore to a much smaller object whose dynamics could, even in principle, be reversed. 

A few comments are in order. 
It could well be that we only want to reverse a particular aspect of the system, and hence do not need to go down to the Planck length and time degrees of precision.
This is fine; our bounds can be adjusted to take into account less precision which will obviously lead to correspondingly larger systems that we could reverse using the universe as a computer. 
Secondly, any inefficiency in our computation (such as using irreversible gates) can only make this task harder. 
Here, we are optimistic in assuming $kT\ln2$ units of energy per bit reset, but this is the ultimate limit of irreversible computation -- current computers are at least five or six orders of magnitude away from this limit~\cite{Frank02,CockshottMM12}.

Finally we note that Niels Bohr emphasized that no phenomenon can truly be a phenomenon until it is irreversibly amplified (i.e.\ recorded in a large enough system)~\cite{PWNB2,WheelerZ83}.
Then, following from the above, no object smaller than ten microns can ever safely be assumed to be a proper quantum measurement apparatus, since it is within the ultimate physical limits of reversibility. 

\vspace*{0.5em}
\textbf{Acknowledgments---}The authors acknowledge funding from the John Templeton Foundation, the National Research Foundation
(Singapore), the Ministry of Education (Singapore), the Engineering and Physical Sciences Research Council (UK), the Leverhulme Trust, the Oxford Martin School, and Wolfson College, University of Oxford. 
This research is also supported by the National Research Foundation, Prime Minister’s Office, Singapore under its Competitive Research Programme (CRP Award No. NRF- CRP14-2014-02) and administered by 
Centre for Quantum Technologies, National University of Singapore.


\begin{thebibliography}{24}
\providecommand{\natexlab}[1]{#1}
\providecommand{\url}[1]{\texttt{#1}}
\expandafter\ifx\csname urlstyle\endcsname\relax
  \providecommand{\doi}[1]{doi: #1}\else
  \providecommand{\doi}{doi: \begingroup \urlstyle{rm}\Url}\fi

\bibitem[Boltzmann(1872)]{Boltzmann1872}
L.~E. Boltzmann.
\newblock {Weitere Studien {\"{u}}ber das W{\"{o}}rmegleichgewicht unter
  Gasmolek{\"{u}}len}.
\newblock \emph{Sitzungsberichte Akademie der Wissenschaften}, 66:\penalty0
  275--370, 1872.
\newblock \doi{10.1142/9781848161337_0015}.

\bibitem[Loschmidt(1876)]{Loschmidt1876}
J.~J. Loschmidt.
\newblock {{\"{U}}ber den Zustand des W{\"{a}}rmegleichgewichtes eines Systems
  von K{\"o}rpern mit R{\"{u}}cksicht auf die Schwerkraft}.
\newblock \emph{Sitzungsberichte der Akademie der Wissenschaften in Wien}, 73,
  1876.

\bibitem[Hahn(1950)]{Hahn50}
E.~Hahn.
\newblock {Spin Echoes}.
\newblock \emph{Physical Review}, 80\penalty0 (4):\penalty0 580--594, Nov 1950.
\newblock ISSN 0031-899X.
\newblock \doi{10.1103/PhysRev.80.580}.
\newblock \url{http://journals.aps.org/pr/abstract/10.1103/PhysRev.80.580}.

\bibitem[Maxwell(1870)]{Maxwell1870}
J.~C. Maxwell.
\newblock \emph{{Theory of Heat}}.
\newblock Longmans, Green {\&} Co, 1902 edition, 1870.

\bibitem[{'t Hooft}(1993)]{tHooft93}
G.~{'t Hooft}.
\newblock {Dimensional Reduction in Quantum Gravity}.
\newblock Oct 1993.
\newblock \url{http://arxiv.org/abs/gr-qc/9310026}.

\bibitem[Susskind(1995)]{Susskind95}
L.~Susskind.
\newblock {The world as a hologram}.
\newblock \emph{Journal of Mathematical Physics}, 36\penalty0 (11):\penalty0
  6377, Nov 1995.
\newblock ISSN 00222488.
\newblock \doi{10.1063/1.531249}.
\newblock \url{http://scitation.aip.org/content/aip/journal/jmp/36/11/10.1063/1.531249}.

\bibitem[Bousso(2002)]{Bousso02}
R.~Bousso.
\newblock {The holographic principle}.
\newblock \emph{Reviews of Modern Physics}, 74\penalty0 (3):\penalty0 825--874,
  Aug 2002.
\newblock ISSN 0034-6861.
\newblock \doi{10.1103/RevModPhys.74.825}.
\newblock \url{http://journals.aps.org/rmp/abstract/10.1103/RevModPhys.74.825}.

\bibitem[Lloyd(2000)]{Lloyd00}
S.~Lloyd.
\newblock {Ultimate physical limits to computation}.
\newblock \emph{Nature}, 406\penalty0 (6799):\penalty0 1047--54, Aug 2000.
\newblock ISSN 1476-4687.
\newblock \doi{10.1038/35023282}.
\newblock \url{http://www.nature.com/nature/journal/v406/n6799/full/4061047a0.html}.

\bibitem[Lloyd(2002)]{Lloyd02}
S.~Lloyd.
\newblock {Computational capacity of the universe.}
\newblock \emph{Physical Review Letters}, 88\penalty0 (23):\penalty0 237901,
  jun 2002.
\newblock ISSN 0031-9007.
\newblock \doi{10.1103/PhysRevLett.88.237901}.
\newblock \url{http://journals.aps.org/prl/abstract/10.1103/PhysRevLett.88.237901}.

\bibitem[Planck(1899)]{Planck1899}
M.~Planck.
\newblock {{\"{U}}ber irreversible Strahlungsvorg{\"{a}}nge}.
\newblock \emph{Sitzungsberichte der K{\"{o}}niglich Preu{\ss}ischen Akademie
  der Wissenschaften zu Berlin}, 5:\penalty0 440--480, 1899.

\bibitem[of~Standards and Technology(2014)]{NIST}
National~Institute of~Standards and Technology.
\newblock {The NIST reference on Constants, Units and Uncertainty}, 2014.
\newblock \url{http://physics.nist.gov/cuu/}.

\bibitem[{Planck Collaboration} Ade et~al.(2015)]{Planck15}
{Planck Collaboration, P.~A.~R.~Ade~et~al.}
\newblock {Planck 2015 results. XIII. Cosmological parameters}.
\newblock Feb 2015.
\newblock \url{http://arxiv.org/abs/1502.01589}.

\bibitem[Valev(2014)]{Valev14}
D.~Valev.
\newblock {Estimations of total mass and energy of the observable universe}.
\newblock \emph{Physics International}, 5\penalty0 (1):\penalty0 15--20, Jan
  2014.
\newblock ISSN 1948-9803.
\newblock \doi{10.3844/pisp.2014.15.20}.
\newblock \url{http://thescipub.com/abstract/10.3844/pisp.2014.15.20}.

\bibitem[Landauer(1961)]{Landauer61}
R.~Landauer.
\newblock {Irreversibility and Heat Generation in the Computer Process}.
\newblock \emph{IBM Journal of Research and Development}, 5:\penalty0 183--191,
  July 1961.
\newblock ISSN 0018-8646.
\newblock \doi{10.1147/rd.53.0183}.

\bibitem[Bennett(1982)]{Bennett82}
C.~H. Bennett.
\newblock {The thermodynamics of computation—a review}.
\newblock \emph{International Journal of Theoretical Physics}, 21\penalty0
  (12):\penalty0 905--940, Dec 1982.
\newblock ISSN 0020-7748.
\newblock \doi{10.1007/BF02084158}.

\bibitem[Leff and Rex(2002)]{LeffR02}
H.~Leff and A.~F. Rex.
\newblock \emph{{Maxwell's Demon 2: Entropy, Classical and Quantum Information,
  Computing}}.
\newblock Taylor {\&} Francis, 2002.

\bibitem[Maruyama et~al.(2009)Maruyama, Nori, and Vedral]{MaruyamaNV09}
K.~Maruyama, F.~Nori, and V.~Vedral.
\newblock {The physics of Maxwell's demon and information}.
\newblock \emph{Rev. Mod. Phys.}, 81\penalty0 (1):\penalty0 1--23, Jan 2009.
\newblock \doi{10.1103/RevModPhys.81.1}.
\newblock \url{http://link.aps.org/doi/10.1103/RevModPhys.81.1}.

\bibitem[B{\'{e}}rut et~al.(2012)B{\'{e}}rut, Arakelyan, Petrosyan, Ciliberto,
  Dillenschneider, and Lutz]{BerutAPCDL12}
A.~B{\'{e}}rut, A.~Arakelyan, A.~Petrosyan, S.~Ciliberto, R.~Dillenschneider,
  and E.~Lutz.
\newblock {Experimental verification of Landauer's principle linking
  information and thermodynamics.}
\newblock \emph{Nature}, 483\penalty0 (7388):\penalty0 187--189, Mar 2012.
\newblock ISSN 1476-4687.
\newblock \doi{10.1038/nature10872}.
\newblock \url{http://www.nature.com/nature/journal/v483/n7388/full/nature10872.html}.

\bibitem[Parrondo et~al.(2015)Parrondo, Horowitz, and Sagawa]{ParrondoHS15}
J.~M.~R. Parrondo, J.~M. Horowitz, and T.~Sagawa.
\newblock {Thermodynamics of information}.
\newblock \emph{Nature Physics}, 11\penalty0 (2):\penalty0 131--139, Feb 2015.
\newblock ISSN 1745-2473.
\newblock \doi{10.1038/nphys3230}.
\newblock \url{http://dx.doi.org/10.1038/nphys3230}.

\bibitem[Fixsen(2009)]{Fixsen09}
D.~J. Fixsen.
\newblock {The Temperature of the Cosmic Microwave Background}.
\newblock \emph{The Astrophysical Journal}, 707\penalty0 (2):\penalty0
  916--920, Dec 2009.
\newblock ISSN 0004-637X.
\newblock \doi{10.1088/0004-637X/707/2/916}.
\newblock \url{http://iopscience.iop.org/article/10.1088/0004-637X/707/2/916}.

\bibitem[Frank(2002)]{Frank02}
M.P. Frank.
\newblock {The physical limits of computing}.
\newblock \emph{Computing in Science {\&} Engineering}, 4\penalty0
  (3):\penalty0 16--26, May 2002.
\newblock ISSN 1521-9615.
\newblock \doi{10.1109/5992.998637}.
\newblock \url{http://dl.acm.org/citation.cfm?id=615614.615933}.

\bibitem[Cockshott et~al.(2012)Cockshott, Mackenzie, and
  Michaelson]{CockshottMM12}
W.~P. Cockshott, L.~M. Mackenzie, and G.~Michaelson.
\newblock \emph{{Computation and Its Limits}}.
\newblock Oxford University Press, 2012.
\newblock ISBN 0199640327.

\bibitem[Bohr(1932--1957)]{PWNB2}
N.~Bohr.
\newblock \emph{{Philosophical Writings of Niels Bohr Vol. 2: Essays 1932--1957}}.
\newblock Ox Bow Press.
\newblock ISBN 0918024536

\bibitem[Wheeler and Zurek(1983)]{WheelerZ83}
J.~A. Wheeler and W.~H. Zurek.
\newblock \emph{{Quantum Theory and Measurement}}.
\newblock Princeton University Press, 1983.
\newblock ISBN 0691613168
\end{thebibliography}
\end{document}